\documentclass[sigplan,nonacm]{acmart}
\usepackage[inline]{enumitem}
\usepackage{graphicx}
\usepackage{listings}
\usepackage[frozencache,cachedir=_minted]{minted}
\usepackage{cleveref}
\usepackage{tikz}

\setminted{%
  fontsize=\footnotesize,
  linenos,
  style=perldoc,
}

\setcopyright{acmlicensed}
\copyrightyear{2018}
\acmYear{2018}
\acmDOI{XXXXXXX.XXXXXXX}

\acmConference[Middleware '25]{26th International Middleware Conference}{December 15--19, 2025}{Nashville, TN}
\acmISBN{978-1-4503-XXXX-X/18/06}

\newcommand\todo[1]{\textcolor{red}{\small \textbf{TODO} #1}}

\begin{document}

\title{Concurrent Scheduling of High-Level Parallel Programs on Multi-GPU Systems}
\subtitle{Preprint, not for Distribution}

\author{Fabian Knorr}
\orcid{0000-0003-4193-374X}
\email{fabian.knorr@uibk.ac.at}
\affiliation{%
  \institution{University of Innsbruck}
  \city{Innsbruck}
  \country{Austria}
}

\author{Philip Salzmann}
\orcid{0000-0002-8668-4639}
\email{philip.salzmann@uibk.ac.at}
\affiliation{%
  \institution{University of Innsbruck}
  \city{Innsbruck}
  \country{Austria}
}

\author{Peter Thoman}
\orcid{0000-0002-4028-7451}
\email{peter.thoman@uibk.ac.at}
\affiliation{%
  \institution{University of Innsbruck}
  \city{Innsbruck}
  \country{Austria}
}

\author{Thomas Fahringer}
\orcid{0000-0003-4293-1228}
\email{thomas.fahringer@uibk.ac.at}

\affiliation{%
  \institution{University of Innsbruck}
  \city{Innsbruck}
  \country{Austria}
}

\renewcommand{\shortauthors}{Knorr et al.}

\begin{abstract}
  Parallel programming models can encourage performance portability by moving the responsibility for work assignment and data distribution from the programmer to a runtime system.
  However, analyzing the resulting implicit memory allocations, coherence operations and their interdependencies can quickly introduce delays into the latency-sensitive execution pipeline of a distributed-memory application.

  In this paper, we show how graph-based intermediate representations help moving such scheduling work out of the critical path.
  In the context of SYCL programs distributed onto accelerator clusters, we introduce the \textit{instruction graph}, a low-level representation that preserves full concurrency between memory management, data transfers, MPI peer-to-peer communication and kernel invocation.

  Through integration within the Celerity runtime, we de\-mon\-strate how instruction-graph scheduling enables a system architecture that performs this analysis concurrently with execution.
  Using a scheduler lookahead mechanism, we further detect changing access patterns to optimize memory allocation in the presence of virtualized buffers.

  We show the effectiveness of our method through strong-scaling benchmarks with multiple Celerity applications on up to 128 GPUs in a production cluster.
\end{abstract}





\maketitle

\section{Introduction}
\label{sec:intro}

Ever since the breakdown of Dennard scaling around 2006~\cite{dennard}, the path of processor innovation is dictated by power constraints and thermal management limits.
In the current era, theoretical compute performance primarily grows with an increase in the number of processing units.
Modern workstations or supercomputers typically present as a hybrid between shared-memory and distributed-memory parallel architectures, with processors integrated through tiered interconnects.
Applications that want to benefit fully from this ongoing architectural shift must be able to effectively distribute data- and task parallel parts of their work onto the ever-growing number of processors within such systems.

Dominant programming paradigms in this space are co-evolving with the hardware landscape.
Being one of the oldest yet still highly relevant programming models, MPI programs work explicitly with the number of participating parallel processors (``ranks'') in order to partition and assign work to ranks and ensure portability between systems.

The more recent but equally popular OpenMP paradigm uses annotations to parallelize conventional constructs like for-loops on shared-memory systems.
It notably does not involve the application developer in work assignment unless they explicitly opt into it through scheduling clauses.
This shifts responsibility to the OpenMP compiler and runtime, allowing newer systems to adapt to existing programs.

When Nvidia's CUDA language made general-purpose GPU computing mainstream, it embraced this philosophy from the start.
The massively-parallel graphics hardware is canonically programmed along an explicitly-parallel SIMT model.
In this framework, the user divides their problem into fine-grained threads to expose all inherent parallelism to the compiler, driver and hardware schedulers, which cooperate to efficiently map the kernel onto hardware resources.
Both the number of processing elements and the processor architecture are hidden from the developer, providing excellent forwards- and backwards compatibility.

Competing APIs on a similar abstraction level, like AMD's HIP or the cross-platform OpenCL and SYCL standards, have followed suit in that regard.
The latter even go a step further by allowing a runtime to take responsibility of memory management by using their \textit{memory object} or \textit{buffer} abstractions.

Extrapolating from this history, we regard automatic and transparent work assignment as the natural evolutionary next step in programming models for hybrid distributed-shared memory parallel systems.
By providing a unified programming interface, these models can continue adapting to hardware topologies of the future without requiring changes in the applications relying on them.

In this paper, we explore how implicit work distribution between GPUs can efficiently be implemented on systems ranging from workstations to large accelerator clusters.

Our core contributions are as follows:
\begin{itemize}
  \item A detailed analysis of the memory management and communication requirements arising from transparent distribution of SIMT programs onto multiple GPUs
  \item The \textit{instruction graph} as an intermediate representation modelling concurrency between these operations
  \item A software architecture based on the Celerity runtime employing instruction-graph scheduling to optimize concurrency and minimize execution latency
  \item A graph lookahead mechanism to dynamically optimize memory allocation patterns on virtualized buffers
  \item Strong-scaling benchmarks on multiple Celerity applications demonstrating superiority of the instruction-graph approach on large GPU clusters.
\end{itemize}

\section{The Celerity Runtime}
\label{sec:celerity}

Celerity~\cite{celerity} is a C++ API and runtime system with a single-source programming model closely following that of SYCL's buffer-accessor model.
The user constructs an out-of-order \textit{queue}, to which they submit a series of \textit{command groups}, which are scheduled to execute as asynchronous \textit{tasks}.
Each task typically launches a single SYCL kernel, declaring reads and writes to data \textit{buffers} using \textit{accessors}.
Accessors carry metadata about the mode and geometry of an access, and act as opaque pointers to device memory within the kernel.

The runtime will compute a parallel schedule for each task by dividing the kernel index space evenly between all SYCL devices in the system.
Data distribution follows work distribution, so responsibility for any memory allocation, coherence copy, communication and synchronization necessary to complete the user program lies with the runtime.

This approach allows existing SYCL programs to be ported to multi-GPU systems with minimal code changes.
In this capacity, Celerity has recently seen use in accelerating SYCL-based physical simulations~\cite{celerity-cronos,celerity-rsim}.

Its high degree of freedom in scheduling has also made the runtime a platform for studying generic dependency-tracking optimizations~\cite{horizons} and automated detection of collective communication patterns~\cite{celerity-gathers}.

We pick Celerity as the basis for our studies in advanced graph-based scheduling techniques because it already operates on high-level graph representation to express work assignment and peer-to-peer transfers between cluster nodes; and its API has the desired characteristic of fully hiding the system topology from the user.

\subsection{Example: $N$-Body Simulation}

\begin{listing}
\begin{minted}{cpp}
using namespace celerity;

const size_t N = 4096;
buffer<double3, 1> P(N); // body positions
buffer<double3, 1> V(N); // body velocities
/* const M, F, T, dt = ... */

queue q;
for (double t = 0; t < T; t += dt) {
  q.submit([&](handler &cgh) {
    accessor p(P, cgh, access::all(), read_only);
    accessor v(V, cgh, access::one_to_one(), read_write);
    cgh.parallel_for<class timestep>(N, [=](item<1> i) {
      double F = 0.0;
      for (size_t j = 0; j < N; ++j)
        F += gravity(p[i], p[j]);
      v[j] += M * F * dt;
    });
  });
  q.submit([&](handler &cgh) {
    accessor v(V, cgh, access::one_to_one(), read_only);
    accessor p(P, cgh, access::one_to_one(), read_write);
    cgh.parallel_for<class update>(N, [=](item<1> i) {
      p[i] += v[i] * dt;
    });
  });
}
\end{minted}
  \caption{%
    A Celerity implementation of direct $N$-Body simulation.
    Each time step submits two tasks (L10, L20) each launching one kernel (L13, L23).
    Their interactions with buffers $P$ and $V$ are mediated through \textit{accessors} that declare access mode and geometry through a \textit{range mapper}.
  }
  \label{lst:nbody}
\end{listing}

\Cref{lst:nbody} shows a simple Celerity implementation of direct $N$-body simulation.
It maintains buffers for the 3-dimensional body positions $P$ and velocities $V$, and in each time step, integrates over the force from all pair-wise gravitational interactions to compute the change in each body's velocity, and the change in position from the updated velocity.

All \textit{work items} (``iterations'') of a \texttt{\small parallel\_for} kernel will execute in parallel from the runtime's point of view, so the program manually resolves the write-after-read / read-after write hazards between updates to $P$ and $V$ by separating the writes into the ``timestep'' and ``update'' tasks.

The kernels' access patterns on $P$ and $V$ are declared through \textit{accessors} constructed with the appropriate access mode \texttt{\small read\_only} and \texttt{\small read\_write}; as well as \textit{range mapper functions} \texttt{\small one\_to\_one} and \texttt{\small all} which describe the relationship between kernel and buffer index space.

With a \texttt{\small one\_to\_one} accessor, kernel and buffer index spaces are identical; while an \texttt{\small all}-accessor always spans the entire buffer range regardless of the kernel.
It is the user's responsibility to ensure that the kernel function adheres to the pattern declared in the range mapper.
We refer the reader to~\cite{celerity} for an in-depth discussion of this concept.


Accessor metadata on every involved buffer is sufficient for Celerity to compute data locality and dataflow resulting from an arbitrary subdivision of work within the cluster.

\subsection{A Case for the Buffer-Accessor Model}
\label{sec:buffer-accessor}

Celerity's buffer-accessor model of memory management originates with SYCL buffers, which in turn inherits their mechanics from opaque OpenCL \textit{memory objects}.
This is in contrast with the explicit pointer-based memory management in CUDA, or newer SYCL standards that support the \textit{unified shared memory} (USM) model.

Pointer-based memory allocation in SYCL requires explicit dependency management, but can reduce kernel register pressure and improve flexibility.
Coupled with some unfortunate API design decisions around SYCL buffers, the pointer-based approach is favored in the single-GPU setting by many contemporary applications~\cite{gromacs-amd}.

In Celerity's scope of distributed and multi-GPU applications however, the buffer-accessor approach reveals its full potential:
Firstly, buffers can be fully virtualized, with the runtime only allocating the parts of the buffer range that is accessed by each GPU.
This allows an application to instantiate and operate on distributed buffers that are many times larger than any single device memory.
Secondly, accessor metadata is a crucial input for the dataflow analysis outlined above, allowing Celerity to present a single coherent global address space (GAS) to the user.

\begin{figure}
  \includegraphics[width=0.9\columnwidth]{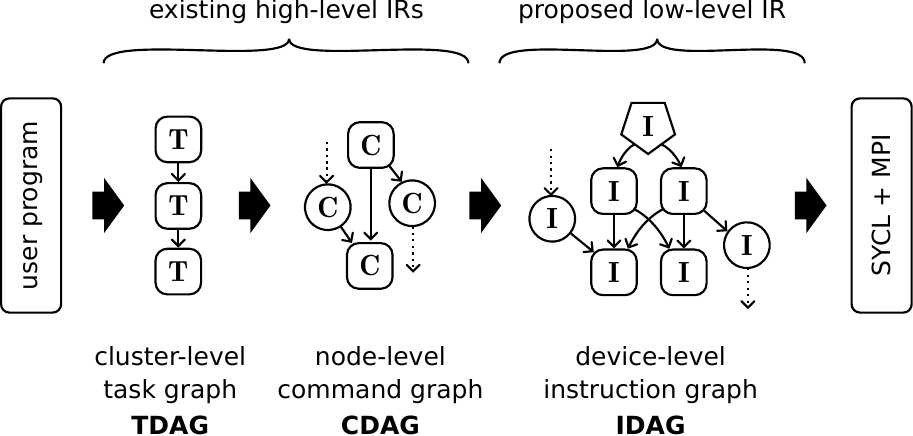}
  \caption{%
    The proposed \textit{instruction graph} complements the established \textit{task-} and \textit{command graph} intermediate representations for distributed GPU programs.
    By modelling individual SYCL and MPI operations, it removes dataflow analysis overhead from the critical execution path.
    }
  \label{fig:dag-overview}
\end{figure}

\subsection{SYCL Graph Semantics}
\label{sec:sycl-dag}

SYCL constrains execution order and concurrency between command groups in terms of an abstract directed acyclic dependency graph (DAG)
\footnote{\url{https://registry.khronos.org/SYCL/specs/sycl-2020/html/sycl-2020.html\#sec:command-groups-exec-order}}.
It mandates a dependency edge between two subsequent command groups when they both access the same buffer and one is a producer access, or when the second one declares an explicit dependency to the first.

SYCL runtimes are then required to execute kernels in an order that fulfills these dependencies.
This permits concurrency, but does not require it:
Notably Intel's DPC++ will serialize execution within out-of-order queues~\cite{aksel-out-of-order}, and the verification-focused SimSYCL does not implement concurrency between SYCL operations at all~\cite{simsycl}.

Celerity follows the spirit of SYCL at the API level, but manages dependencies at a granularity of individual buffer elements rather then entire buffers with the help of range mappers.
As we will see below, graph structures are effective as more than a mere specification tool.
Employed as intermediate scheduling representations, they help expose concurrency between explicit and implicit operations.

\subsection{State of the Art: Command-Graph Scheduling}
\label{sec:tdag-cdag}

Celerity today\footnote{\url{https://github.com/celerity/celerity-runtime/releases/tag/v0.5.0}} generates two graph-based intermediate representations from the sequence of user command group submissions to divide work between cluster nodes:

Each \textit{task} in the \textit{task graph} (TDAG) represents an operation the cluster will execute collectively, most commonly a compute kernel.
The task graph is generated identically on all nodes, and its dependencies are computed as if it the program were executing on a single device.

From the task graph, all nodes collectively generate the \textit{command graph} (CDAG), which distributes the task kernel index space onto nodes and models the peer-to-peer communication necessary to satisfy the resulting data dependencies.
As a distributed process, each node only generates the specific part of the command graph it will itself later execute.
This is a central design decision which keeps Celerity scheduling scalable to a large number of nodes~\cite{celerity-ccgrid}.

\begin{figure}[h]
  \includegraphics[scale=0.42]{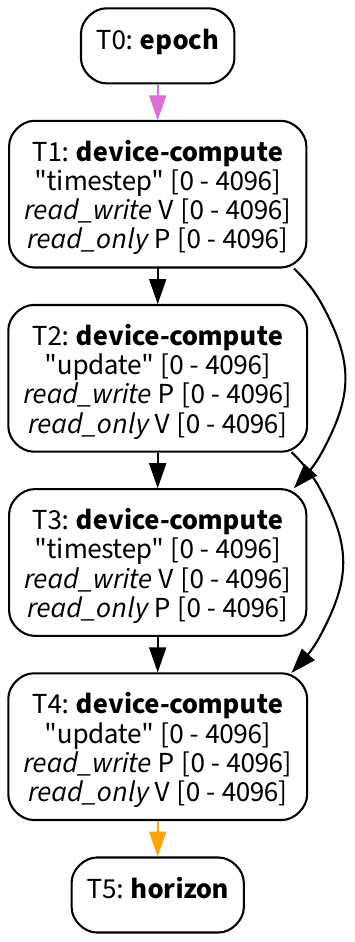}
  \hfil
  \includegraphics[scale=0.42]{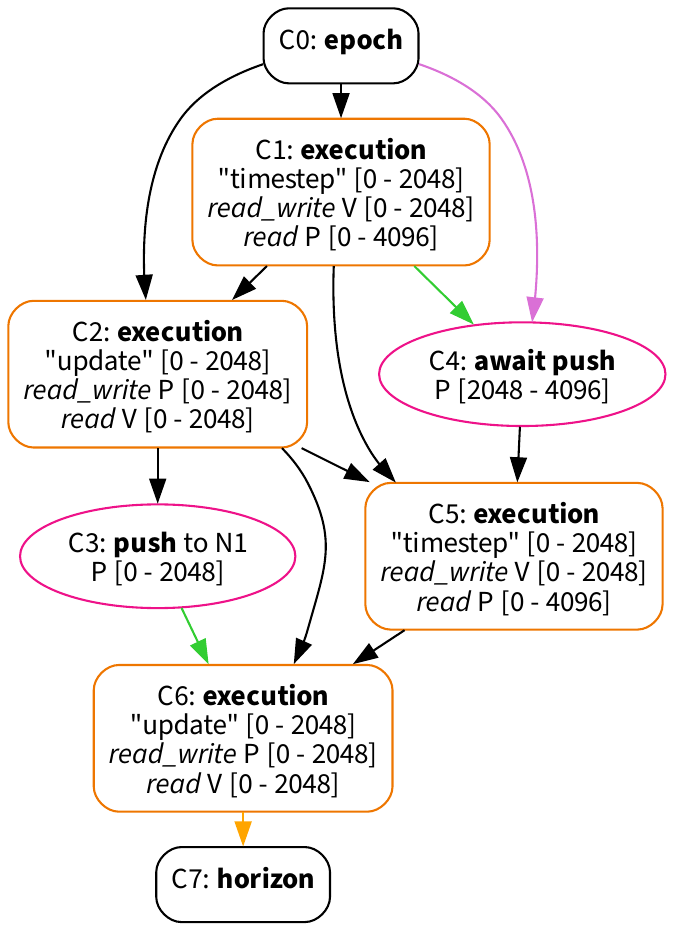}
  \caption{%
    Task graph (left) and command graph (right) computed by node $N0$ out of 2.
    Dataflow dependencies are colored black, anti- and output dependencies green, and graph-synchronization dependencies violet or orange.
  }
  \label{fig:tdag}
\end{figure}

\Cref{fig:dag-overview} visualizes the relationship between TDAG and CDAG, and illustrates how the instruction graph we will introduce in \cref{sec:idag} fits into this picture.
\Cref{fig:tdag} shows the task graph and partial command graph that Celerity generates on node \textit{N0} (i.e.\ MPI rank 0) from the first two loop iterations in \cref{lst:nbody}, when running on two nodes.

For the simplicity of the example shown, the initial values of $P$ and $V$ are assumed to already reside on each node when the buffers are created.
Each \textit{task} reads or writes the full range of buffers $P$ and $V$, since task and buffer size match in the code, their index spaces are equal.
From this access pattern, the runtime infers a linear dependency chain.

At predefined locations, Celerity will insert \textit{epoch} tasks, which are a means of graph-based synchronization~\cite{epochs}, and \textit{horizon} tasks, which bound scheduling complexity by periodically compacting tracking structures~\cite{horizons}.

The partial command graph on node \textit{N0} operates on the first half of the task index space, following a static work assignment scheme.
To satisfy the \texttt{\small all}-read in the second ``timestep'' kernel \textit{C5}, \textit{N0} will \textit{await} receiving the remaining range of $P$ that has been produced in parallel by its peer during \textit{T2}.
In exchange, it will \textit{push} its half of $P$ to its peer node as soon as it the data is available from \textit{C2}.

Dependencies in the CDAG are of a finer granularity than those in the TDAG:
Edges between tasks are replaced by explicit data transfer commands where necessary, and these can expose additional concurrency.
In our example, \textit{C4} originates from \textit{T3}, but may execute concurrently with \textit{C2}, which in turn originates from \textit{T2} -- even though \textit{T2} is a predecessor of \textit{T3}.
This construction allows Celerity to overlap computation in one task with communication from another, helping to hide communication latency.

\subsection{Legacy: Ad-Hoc Memory Management}
\label{sec:buffer-manager}

Celerity, without the advancements proposed in this paper, will forward the generated stream of commands directly to its \textit{executor} state-machine.
This component tracks the progress of each command, and triggers execution of its successors in the dependency graph once they become ready.
The submitted SYCL kernels and MPI operations are allowed to complete asynchronously, meaning that independent commands can indeed execute concurrently as desired.

Because the command graph is oblivious to memory management and local buffer coherence, the executor must perform additional bookkeeping to dispatch these memory operations on the fly.
While the constituent SYCL operations can overlap with unrelated commands just like the kernel itself, they must still execute as a single indivisible sequence.
This forfeits potential for local computation-communication overlap between copy- and kernel operations.

Additionally, the dataflow analysis required for ad-hoc buffer coherence becomes part of the executor's critical path, introducing latencies that can hurt overall scalability.

As a final downside, executing commands immediately as they are generated, and thus allocating buffer memory on first touch, means that a sequence of tasks with varying access patterns may trigger costly allocations resizes.

We will resolve all these shortcomings of the baseline Celerity version by introducing the \textit{instruction graph} as a new low-level intermediate representation.

\section{Instruction-Graph Scheduling}
\label{sec:idag}

We first recognize that commands decompose into local SYCL and MPI operations as discussed in \cref{sec:buffer-manager}, and that this schedule of kernel launches, data transfers an memory allocations can itself be expressed as a dependency graph.

By constructing a single graph encompassing the constituent operations from multiple independent commands, or the data-parallel assignment of a command kernel index space onto multiple GPUs, we can expose the full concurrency inherent in the execution of a program.

We name these local micro-operations \textit{instructions} and propose the \textit{instruction graph} (IDAG) as a third, low-level intermediate graph representation in Celerity scheduling.
Following \cref{fig:dag-overview}, the instruction graph can be generated directly from the command graph, either ahead-of-time or concurrently with the execution of earlier tasks.

Instructions exist at the abstraction level of the individual SYCL and MPI operations, but also include a few synchronization and lifetime management primitives.
\Cref{tab:instructions} lists all instruction types relevant to this paper that we propose for multi-GPU scheduling.
Our implementation additionally covers instructions to manage host objects, fences, and reductions to recreate the full feature set of the baseline runtime, but we consider them out of scope for this discussion.

It should be noted that the instruction graph is different from a SYCL DAG (see \cref{sec:sycl-dag}), since the latter has no notion of memory management operations or the host-side MPI- and synchronization work Celerity must perform.

\begin{table}[h]
  \centering
  \small\sffamily
  \smallskip
  \renewcommand{\arraystretch}{0.95}
  \begin{tabular}{ll}
    \toprule
    \textit{alloc} & allocate host or device memory \\
    \textit{copy} & 1/2/3D copy between allocations \\
    \textit{free} & free host or device memory \\
    \midrule
    \textit{send} & perform an \texttt{\small MPI\_Isend}\\
    \textit{receive} & perform one or more \texttt{\small MPI\_Irecv}s\\
    \textit{split receive} & initiate receive with consumer split\\
    \textit{await receive} & await subregion of a \textit{split receive}\\
    \midrule
    \textit{device kernel} & launch SYCL kernel on device\\
    \textit{host task} & launch host task functor in thread \\
    \midrule
    \textit{horizon} & prune graphs in scheduler\\
    \textit{epoch} & synchronize with main thread\\
    \bottomrule
  \end{tabular}
  \bigskip
  \caption{%
    Instruction types proposed for Celerity IDAG scheduling; grouped by memory management, peer-to-peer communication, compute, and synchronization instructions.
  }
  \vspace*{-1em}
  \label{tab:instructions}
\end{table}


\subsection{Hierarchical Work Assignment}
\label{sec:work-assignment}

As part of CDAG generation, Celerity distributes work between cluster nodes by statically splitting the task kernel index space along one or more axes.
Parameters of the split can be user-controlled through the \textit{hint} and \textit{constraint} APIs~\cite{celerity-ccgrid}.
One kernel execution command is emitted per node, requiring the user to spawn one MPI rank per device.

For instruction-graph scheduling, we need to manage all devices of a multi-GPU node within a single process to maximize local concurrency.
To distribute work between local devices, we split the command kernel index space by applying the above scheme a second time.

For large enough task index spaces, this results in one \textit{device kernel} instruction per device (or \textit{host-task} instruction per node), all of which are fully concurrent initially.

\subsection{Memory Allocation}
\label{sec:idag-memory}

Buffer virtualization is one of the core features of Celerity, enabling weak-scaling to problem sizes that far exceed the memory capacity of any single device.
Correctly handling all instances of varying buffer access patterns is therefore a core responsibility of the IDAG generator.

The instruction graph operates in terms of \textit{allocations}, which can either back a buffer subregion on the device or host, or be used as scratch memory in various circumstances.
They are used as the source and destination of all other instructions that operate on runtime memory.
Within kernels and host tasks, allocation pointers are interpolated into accessors before the kernel or host function executes.

Since concrete memory addresses only become known once the corresponding \textit{alloc} instruction has completed, they are represented in the graph by numeric \textit{allocation ids}.

GPU-accelerated systems typically have multiple disjoint hardware memories between which data must be transferred explicitly.
Each allocation is thus associated with a \textit{memory id} that identifies the memory space it resides on.

We enumerate them as \textit{M0} for user-controlled host memory, \textit{M1} for DMA-capable, page-locked host memory, and \textit{M2} or higher for dedicated device memories.
Typical HPC systems will have a 1:1 mapping between \textit{device ids} \textit{D0, D1, \ldots} and memory ids \textit{M2, M3, \ldots}, but this construction can support systems that share physical memory between host and device, or between multiple devices, as well.

Unlike ad-hoc memory management in baseline Celerity, the IDAG permits multiple non-overlapping backing allocations to coexist per buffer.
This allows us to support non-rectangular access patterns
without having to allocate unused elements in the buffer index space bounding box.

To limit complexity however, we require that buffer backing allocations remain non-overlapping, and that each accessor within a kernel or host task is backed by a single contiguous allocation.
This might require us to perform allocation resizes, i.e.\ a chain of \textit{alloc}, \textit{copy} and \textit{free} operations to extend or merge existing allocations when the access patterns change during the lifetime of a buffer.

\begin{figure}[h]
  \includegraphics[width=0.95\columnwidth]{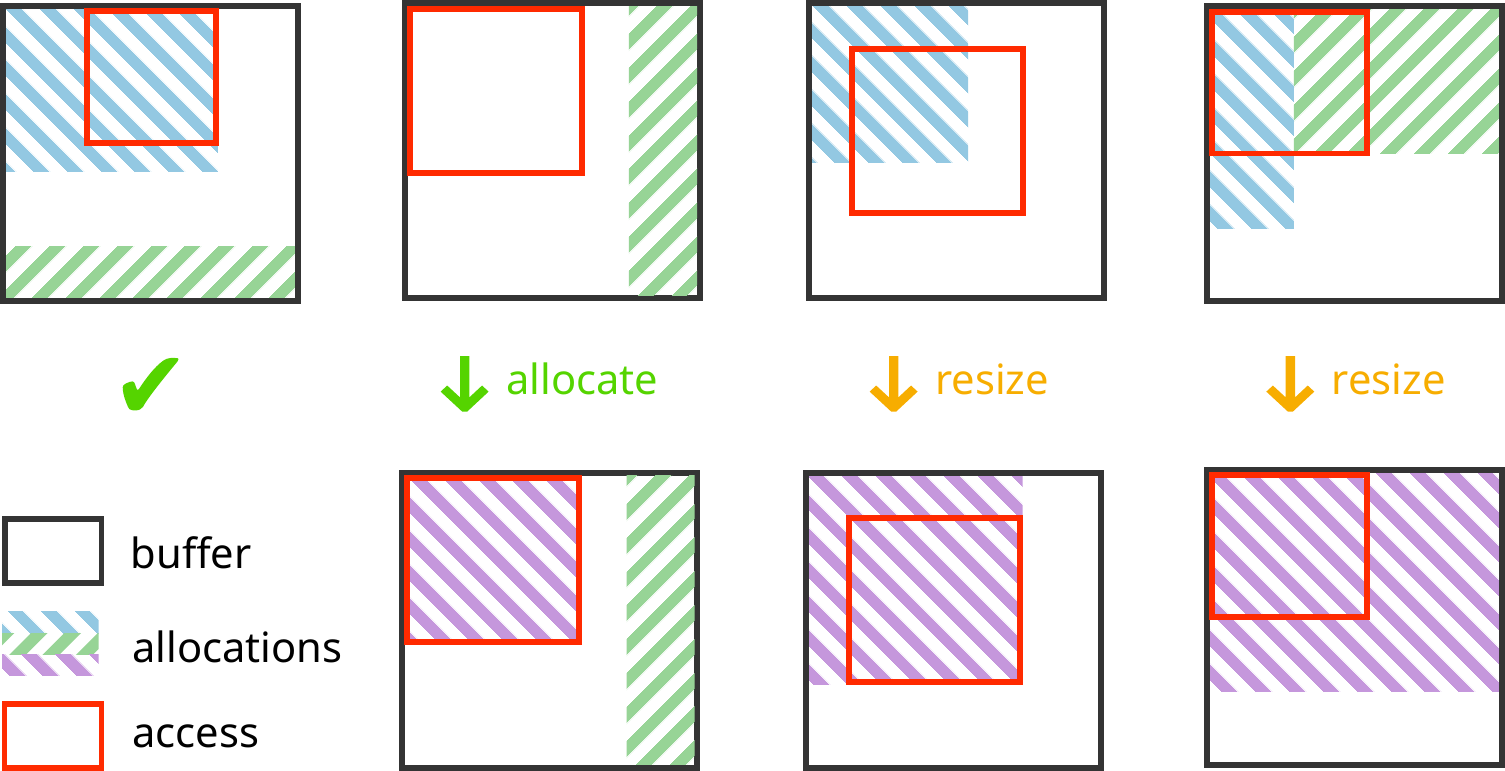}
  \caption{%
    Because accessors map to a single device-memory pointer, each must be backed by a contiguous allocation.
    Kernel launches will be preceded by the necessary allocation- and resize-copy instructions in the instruction graph.
  }
  \label{fig:buffer-alloc}
\end{figure}

\Cref{fig:buffer-alloc} visualizes all situations that may occur when the necessary allocations for a kernel launch are materialized.

Note that we never emit instructions to downsize a buffer allocation, as we do not know which subregion will be required again in the future.
All allocations are however returned back to the system eventually by emitting appropriate \textit{free} instructions.
For buffer backing allocations, this happens as soon as the user drops their last reference to the buffer, and the last task accessing that buffer has completed.

\subsection{Local Buffer Coherence}
\label{sec:coherence}

Before launching a kernel that reads from a buffer, the accessed backing allocations must be made coherent with the most recent buffer version.
By keeping track of which buffer subregions are locally up-to-date on what memory ids, and which instruction has been its local \textit{original producer}, we can insert the appropriate \textit{copy} instructions into the graph.

When input data has been computed on a different node, a previous \textit{await push} command will have made the affected subregion coherent in host memory, and we consider the associated \textit{receive} instruction to be its local original producer.

To retain maximum concurrency, any copied region is subject to \textit{producer-} and \textit{consumer split}:
Instead of generating as few copy instructions as possible, we insert one copy for each pairing of original-producer and consumer instruction.
This ensures that copies do not create artificial synchronization points in the graph, and that subregions available early can be copied to the target memory right away to further facilitate computation-communication overlap.

On multi-GPU systems that do not support device-to-device copies in hardware (as is true for most contemporary consumer GPUs), we emit appropriate sub-graphs of copy instructions to stage the data in host memory.

\subsection{Peer-to-Peer Communication}
\label{sec:idag-p2p}

Using a simplified form of the coherence tracking outlined above, Celerity decides during CDAG generation which buffer regions need to be exchanged with peer nodes in the form of \textit{push} and \textit{await-push} commands.

These commands are asynchronous in nature:
While \textit{push} commands carry the receiver node together with the precise buffer region to be transmitted,
their \textit{await push} counterparts only know about union of all subregions that will be received as part of the task, and hold no information about the participating sender nodes.

This peculiarity originates from a trade-off in distributed command generation which bounds tracking complexity to ensure Celerity's scalability to large clusters~\cite{celerity-ccgrid}, and has far-reaching consequences for the way receiving data must be handled during scheduling and execution.


\subsubsection*{Outbound Transfers}

For each \textit{push} command, we trigger a coherence copy to host memory and generate a \textit{send} instruction for every rectangular region to be transmitted; again subject to producer split.

Each \textit{send} is tagged with a locally unique \textit{message id} that will be used to match send- and receive instructions between peers at execution time.
The association between message id and sent region is transmitted to the receiver ahead of execution time in form of a \textit{pilot message}.

\subsubsection*{Inbound Transfers}

Since no information on senders is contained in \textit{await push} commands, instruction graph generation for inbound transfers is similarly constrained.

Thus, when all received data is to be consumed by a single instruction, or all dependents consume the entire awaited region, we emit a single \textit{receive} instruction that will place the incoming data into host memory.
Ingesting pilot messages to dispatch the actual MPI operations is left to the executor.

Applying a consumer split in the case of multiple disjoint or overlapping consumers is more complex than with local coherence copies.
Since the sender geometry is unknown, one of three situations will occur at execution time:
\begin{enumerate}[label=\alph*)]
\item\label{items:inbound:a} Multiple senders transmit subregions in the exact geometry we consume.
  This is the ideal path and provides optimal computation-communication overlap.
\item\label{items:inbound:b} A single sender satisfies the entire \textit{await push} region at once.
  This renders consumer-split ineffective and requires us to be able to receive the entire region into contiguous host memory.
\item\label{items:inbound:c} Multiple senders transmit disjoint subregions, but in a geometry orthogonal to the receiver split.
  This might allow some consumer instructions to begin executing early, but might also degrade to case \ref{items:inbound:b}.
\end{enumerate}

The instruction graph must ensure a contiguous backing allocation for the entire \textit{await push} region to prepare for option \ref{items:inbound:b}.
To profit from either \ref{items:inbound:a} or \ref{items:inbound:c}, it will then emit a \textit{split receive} instruction carrying the target host allocation id, followed by multiple \textit{await receive} instructions subject to consumer split.

The executor is then tasked with recognizing any \textit{await receive} as completed as soon as its subregion or a superset thereof has been received, regardless of the geometry of inbound transfers that satisfied the request.


\subsection{Synchronization Instructions}
\label{sec:idag-sync}

During task graph generation, Celerity emits \textit{horizon tasks} based on critical-path length and the number of parallel independent tasks.
These effectively bound scheduling complexity by limiting the set of tasks or commands that are considered \textit{original producers} of any buffer region~\cite{horizons}.
We apply this same principle to the instruction graph as well.

Executing a horizon instruction also functions as a forward-progress indicator, which we use to periodically prune parts of the IDAG that have finished and are no longer referenced.

As we will see in \cref{sec:lookahead}, the horizon mechanism also allows us to establish a lookahead window on the command graph without walking dependencies.
This scheduler lookahead will allow us to avoid committing to inefficient buffer backing allocations too early.

\subsection{Instruction Graph for the $N$-Body Example}
\label{sec:idag-nbody}

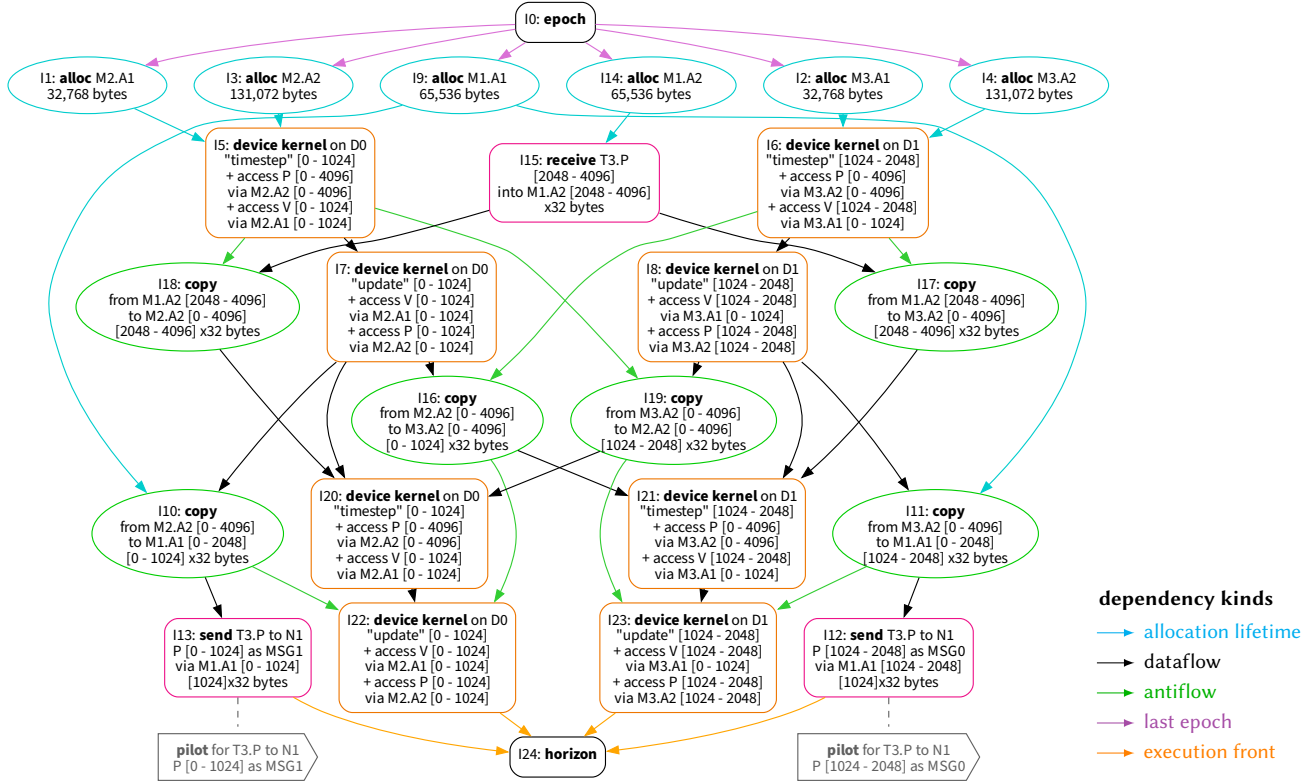
\begin{figure*}[h]
  \hfill
  \includegraphics[scale=0.42]{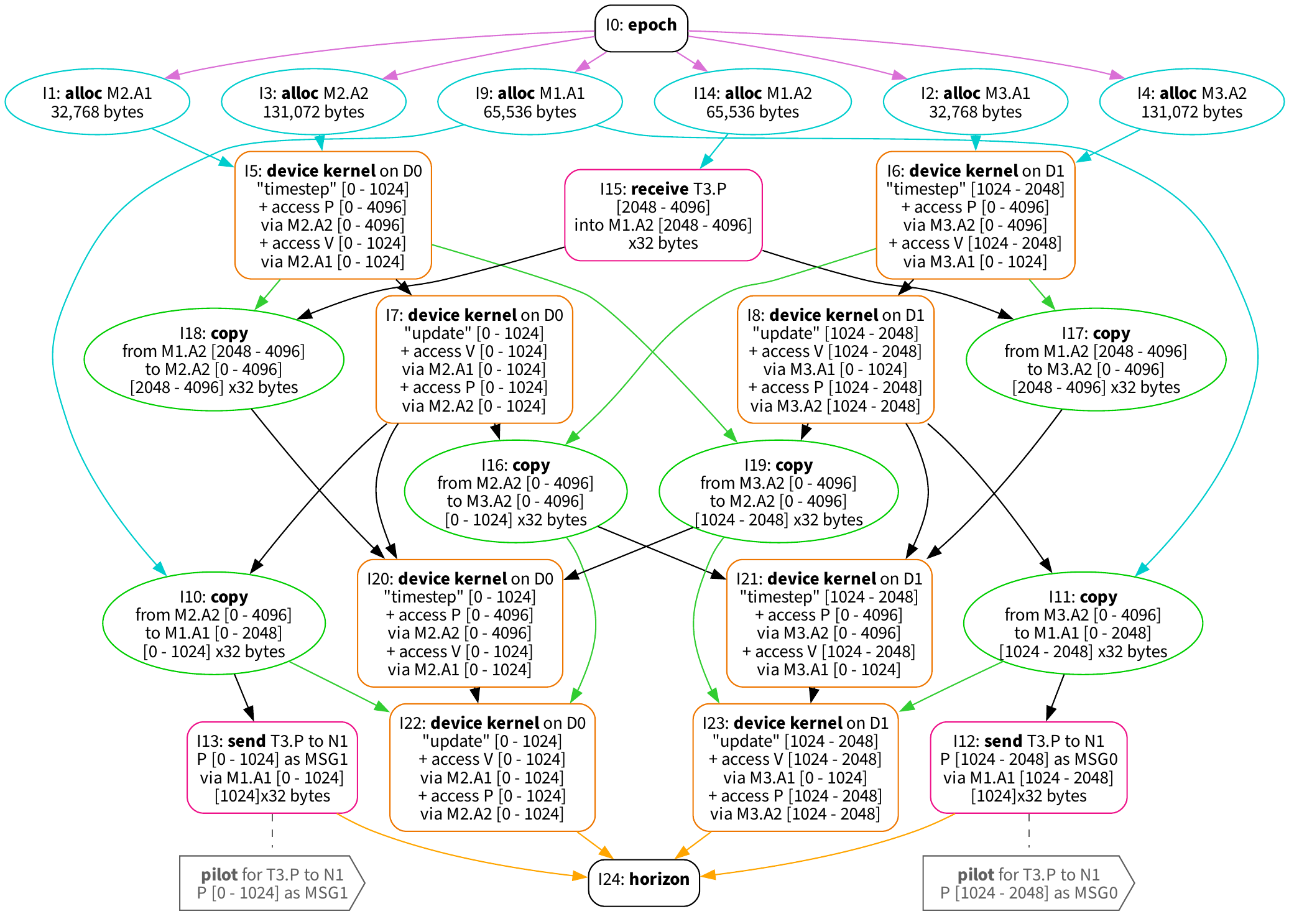}
  \hspace{-5mm}
  \begin{tikzpicture}[font=\sffamily\footnotesize,xscale=0.5,yscale=-0.4]
    \draw (-0.2,-0.1) node[right]{\textbf{dependency kinds\vphantom{bp}}};
    \draw[-latex,color=cyan] (0,1) -- (1,1) node[right] {allocation lifetime\vphantom{bp}};
    \draw[-latex,color=black] (0,2) -- (1,2) node[right] {dataflow\vphantom{bp}};
    \draw[-latex,color=green!70!black] (0,3) -- (1,3) node[right] {antiflow\vphantom{bp}};
    \draw[-latex,color=violet!70!white] (0,4) -- (1,4) node[right] {last epoch\vphantom{bp}};
    \draw[-latex,color=orange] (0,5) -- (1,5) node[right] {execution front\vphantom{bp}};
    \draw (0,6) node{}; 
  \end{tikzpicture}
  \hfill
  \vspace{-2mm}

  \caption{%
    Instruction graph compiled from the command graph in \cref{fig:tdag} for the two local devices \textit{D0} and \textit{D1} on node \textit{N0}.
    Each device receives half the command kernel index space (a quarter of the task kernel index space).
    Note that all \textit{send}- and \textit{receive} instructions in this picture are concurrent, even though they appear far apart due to layout constraints.
  }
  \label{fig:idag}
\end{figure*}

\Cref{fig:idag} shows the instruction graph for the $N$-body example in \cref{lst:nbody}, as it is generated for two devices on node \textit{N0}.
Using the default one-dimensional split, the scheduler assigns the first half of the command kernel index space (i.e.\ the first \textit{quarter} of the task kernel index space) to device \textit{D0}, and the second half (quarter) to device \textit{D1}.

Before the first ``timestep'' kernel, we emit \textit{alloc} instructions to create backing allocations for the the entirety of buffer $P$ on $D0$'s native memory $M2$ and $D1$'s native memory $M3$, as required by the \texttt{\small all} range mapper.
For the read-write access on $V$, the \textit{\small one\_to\_one} mapper declares that buffer and kernel index space match, meaning the first quarter of buffer $V$ must be allocated in $M2$, and second quarter in $M3$.

We then launch the first pair of ``timestep'' (\textit{I5}, \textit{I6}) and ``update'' kernels (\textit{I7}, \textit{I8}) on each device.
Between the kernels on one device, all data dependencies are trivially fulfilled by them operating on the same buffer allocations.

The second kernel is followed by a pair of \textit{push} and \textit{await push} commands on buffer $P$ in the CDAG.
The push command \textit{C3} is compiled to a pair of \textit{send} instructions \textit{I10} and \textit{I11} following the producer-split logic.
Each \textit{send} instruction is preceded by a device-to-host copy to make the input data available to MPI.
Both \textit{send}s are accompanied by pilot messages that are transmitted to the peer \textit{N1} immediately.

Likewise, the \textit{await push} command \textit{C4} is compiled to a single \textit{receive} instruction.
The consumer-split logic does not apply, as both consumers require the same buffer region.

On the second ``timestep'', both device kernels will again consume the buffer $P$ which has now been updated, in part on each device, and in part on the host by the \textit{receive} instruction.
We initiate copy instructions to make the entirety of $P$ coherent on both $M2$ and $M3$ from all three sources.
Between the two local devices, this results in a pair of device-to-device copies \textit{I16} and \textit{I17}, which can begin executing as soon as the kernel producing that data has completed.

Once buffer coherence is established, the kernels of the second iteration can launch.
Afterwards, the \textit{horizon} command \textit{C7} is compiled into \textit{horizon} instruction \textit{I24}, which by definition depends on all instructions on the current execution front, i.e.\ those instructions without another successor.

The instruction graph for the original program will continue repeating the graph topology between instructions \textit{I10} through \textit{I24} until the loop in \cref{lst:nbody} exits.

The dependency relationships in \cref{fig:idag} highlight how operations that would be serialized in the CDAG can partially overlap in the IDAG and perform coherence copies and peer-to-peer communication concurrently with kernel execution.

\section{Proposed System Architecture}
\label{sec:architecture}

\begin{figure*}[h]
  \centering
  \includegraphics[width=\linewidth]{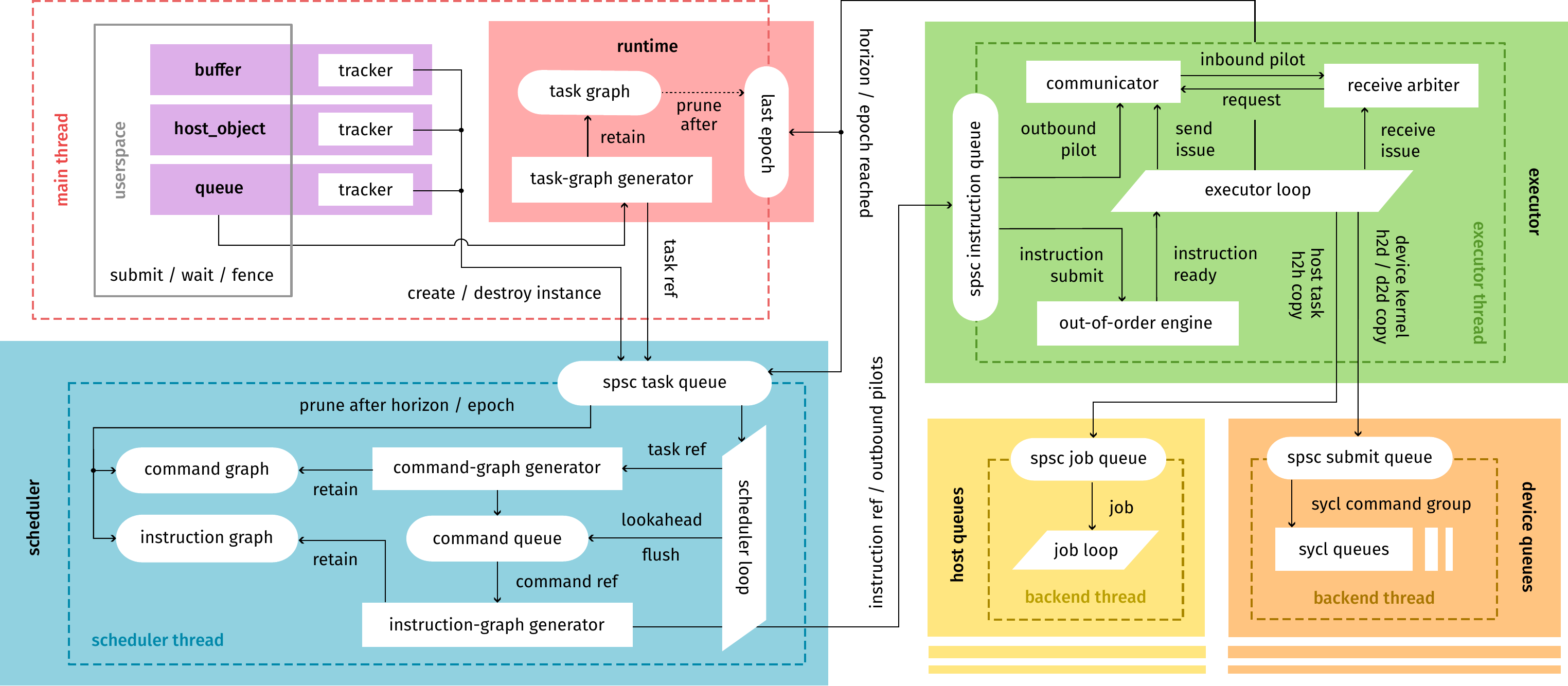}
  \caption{%
    Proposed concurrent architecture of Celerity with instruction-graph scheduling.
    The user-controlled main thread, graph scheduler, executor state machines and backends are all decoupled to operate concurrently and communicate over single-producer-single-consumer (spsc) queues.
    Dashed lines represent thread boundaries.}
  \label{fig:architecture}
\end{figure*}

A central advantage of the instruction-graph approach is its capacity to decouple the generation of a program schedule from its execution.
This allows scheduling of new tasks to occur concurrently with the execution of older ones, as well as asynchronously to the user program.
We propose the architecture outlined in \cref{fig:architecture} to fully exploit this capability.

Next to the user-controlled main thread, we introduce a dedicated scheduler thread for combined command- and instruction graph generation, as well as an executor thread harboring the state machines necessary to drive instructions to completion.
The executor in turn offloads the submission of host and device work to separate backend threads in order to remove their latency from its main loop.

Instead of operating on shared data structures that would require synchronization, all inter-thread communication is unidirectional and mediated by single-producer-single-consumer (spsc) queues, decoupling the individual components to minimize wait states.
References to tasks, commands and instructions are guaranteed to remain live during execution via the horizon mechanism (\cref{sec:idag-sync}).

\subsection{Out-of-Order Instruction Dispatch}
\label{sec:out-of-order}

The IDAG scheduler generates instructions in topological order, i.e.\ a sequential execution of the instruction stream would trivially fulfill all internal dependencies.
To benefit from parallel hardware resources especially in a multi-GPU scenario however, multiple instructions must be kept in-flight concurrently whenever their dependency relationship allows it.
To this end we dynamically allocate multiple SYCL in-order queues per device for concurrent device-side copy and kernel instructions, as well as a series of host threads for host tasks and host-side copy instructions.

Submitting concurrent operations to in-order queues and polling them for asynchronous completion has previously been shown to deliver excellent overlapping capabilities for complex dependency graphs in CUDA applications~\cite{cuda-middleware}.
A more explicit polling-based approach has been used in the Charm++ runtime to invoke host work depending on asynchronous CUDA operations~\cite{charmpp-overdecomposition}, where the authors discuss the trade-off between the event polling and the alternative of CUDA stream callbacks.

In this spirit, we propose an executor loop issuing ready instructions and polling active ones for completion.
Strong-scaling behavior, i.e.\ the performance impact of submitting ever shorter-running instructions, is highly sensitive to latency in both instruction selection and polling, so as little time as possible must be spent in either.

We propose the \textit{out-of-order engine} state machine to handle both instruction selection and retirement.
It is fed the stream of incoming instructions as well as completion events, and will select the next instruction to be issued to a backend queue.
An instruction can either be assigned \textit{directly} when all its dependencies are satisfied; or \textit{eagerly} when all its incomplete dependencies are currently pending on the same single in-order queue or host thread, exploiting the fact that the backend's FIFO semantics will implicitly guarantee the correct dependency ordering.

\subsection{Peer-to-Peer Receive Arbitration}
\label{sec:recv-arbitration}

Within the IDAG, \textit{send} instructions carry their target node id and the precise buffer region to be transferred.
As such, they are passed directly to the \textit{communicator} abstraction, which issues an appropriate \texttt{\small MPI\_Isend} for the transfer.

However, as a side-effect of distributed command-graph generation, \textit{receive} instructions only carry the union of buffer regions that will be received within a task as outlined in \cref{sec:idag-p2p}.
The information which peer nodes will contribute what subregion, necessary to issue a matching \texttt{\small MPI\_Irecv}, becomes known early at execution time in the form of \textit{pilot messages} transmitted by each contributing sender.

We call the process of matching \textit{receive} and \textit{split receive} instructions to pilots in order to precisely compute the source and geometry of incoming transfers \textit{receive arbitration}.
We implement it using a state machine which places \texttt{\small MPI\_Irecv} calls on the destination memory through the communicator as soon as source and geometry of the transfer are known.

Since \textit{receive} instructions cannot have local dataflow dependencies, calls to \texttt{\small MPI\_Irecv} can typically be issued long before the sender side begins transmitting, helping to eliminate any implicit double-buffering otherwise done by MPI.

\subsection{Scheduler Lookahead and Resize Elision}
\label{sec:lookahead}

Memory allocations in GPU programs are typically very slow, both for device memory and DMA-capable (page-locked) host memory.
Drivers will map virtual pages to physical memory immediately to avoid complexities like page-fault handling.
This is acceptable in most GPU programs which allocate memory for inputs and outputs once ahead of time, but read and write to them repeatedly.

Celerity however, with its unique combination of concurrent scheduling and virtualized buffers, cannot know the full buffer subregion that each device will access over the lifetime of a program.
A backing allocation that is made according to the requirements for an earlier task may be insufficient for a later one, causing a chain of \textit{alloc}, \textit{copy} and \textit{free} instructions to resize it (see \cref{fig:buffer-alloc} and \cref{lst:resize}).

Not only do these resizes incur a performance penalty---programs might in fact run out of memory when the combined allocation sizes exceed available device memory; a scenario that is likely in weak-scaling applications.

\begin{listing}[h]
    \medskip
\begin{minted}{cpp}
buffer<float, 2> buf(...);
q.submit([&](handler &cgh) {
  accessor write_interior(buf, cgh,
           access::one_to_one(), write_only, no_init);
  cgh.parallel_for(buf.get_range(), ...);
});
q.submit([&](handler &cgh) {
  accessor read_neighborhood(buf, cgh,
           access::neighborhood({1, 1}), read_only);
  cgh.parallel_for(buf.get_range(), ...);
});
\end{minted}
\caption{%
  A task writing with a one-to-one mapping, followed by a read of the one-neighborhood of the same region.
  Scheduling this sequence naively will cause a resize allocation in the second task due to the growing access pattern.
}
\label{lst:resize}
\end{listing}

\subsubsection*{Command Queueing}

We first recognize that the need for a resize can be identified early by analyzing a larger sequence of tasks instead of generating instructions for each task in isolation.
In fact, a program in which the user never explicitly synchronizes with the runtime could be fully scheduled ahead-of-time, making the precise memory requirements known before the first allocation is made.
This however would have the obvious downside of losing much if not all scheduler concurrency.

Instead, we propose a lookahead mechanism that will generate task- and command graph immediately, but will heuristically postpone instruction-graph generation as long as changing memory allocation patterns can be observed.
Whenever a new command has been generated, the scheduler will inquire whether compiling it right away would emit any \textit{alloc} instructions, and mark the command as \textit{allocating} if so.
Recognizing this condition is inexpensive compared to generation of the actual instruction graph.

We then place each command into the scheduler's \textit{command queue} (\cref{fig:architecture}).
As long as no allocating command is queued, we can dequeue and compile any commands immediately.
Once the first allocating command enters the queue, we stop generating instructions, expecting that another allocating command will follow soon with which we can merge allocations to avoid a potential resize.

\subsubsection*{Lookahead Heuristic}

To avoid losing concurrency when buffering commands too far into the future, we flush the command queue once we have encountered two horizons (see \cref{sec:idag-sync}) after the last allocating command.

Due to how horizons are generated in the TDAG, this condition is indicative of a task chain accessing the same buffers repeatedly in compatible patterns.
The arbitrary but deterministic choice of graph distance correctly predicts many common patterns where requirements change within the first few tasks, but enter a ``steady state'' eventually.

When we finally generate the instruction graph for the first allocating command, we extend the region covered by each \textit{alloc} instruction to cover all requirements we observed while the command was in the queue.
This way, the requirements of all subsequent commands will already be fulfilled by the time their instruction graph is generated.

\Cref{sec:experiments} will showcase RSim, an application for which memory requirements grow by one buffer row each time step.
For this pattern, the horizon-based heuristic will never flush the command queue, causing the entire command graph to be generated before the first instruction is emitted.
This successfully eliminates any resizes, but loses all concurrency between CDAG generation and execution.
IDAG generation however, which thankfully constitutes the bulk of scheduling work, can still proceed concurrently.

\subsection{User-Facing Debug Facilities}
\label{sec:debug}

Celerity's range-mapper based access model allows us to verify that the user program is behaving according to its specification in various situations.
We introduce a selection of runtime correctness checks that are conditionally enabled in debug builds to aid in verification.

\subsubsection*{Uninitialized-Read Detection}
The contents of a newly created SYCL or Celerity buffer is uninitialized by definition, so no correct program must immediately read from it.
This implies that whenever a reading access is declared on a buffer, the consumed region must either have been initialized explicitly or written by a previous task.

We therefore track the union of all initialized and written-to regions per buffer, and emit a scheduler warning when a reading accessor is constructed on uninitialized data.

\subsubsection*{Overlapping-Write Detection}

As part of its implicit buffer coherence mechanism, Celerity enforces that when a task is split into multiple concurrent instructions, each must write to disjoint buffer regions.
This allows the scheduler to track which node and which instruction is the original producer of any buffer item, and emit the appropriate coherence operations when region is read by a third device.

When the user program violates this assumption, for example by constructing a writing accessor with an \texttt{\small all} range mapper, buffer coherence can no longer be established reliably.
To diagnose this, we detect whenever the regions written by multiple nodes or devices overlap and emit a scheduler error message.

\subsubsection*{Accessor Bounds Checking}

Since Celerity accessors, like their SYCL counterparts, grant access to buffer memory via the array subscript operator, we are able to detect when any individual read or writes fall outside the subregion permitted by the range mapper.
Unless the user performs explicit pointer arithmetic, we are able record the element indices involved in any out-of-bounds access within a kernel, and will report their bounding box in a runtime error message after the kernel exits.


\section{Experiments}
\label{sec:experiments}

While this is an architecture paper first and foremost, we want to demonstrate the benefits of instruction-graph scheduling and the resulting system architecture in comparison with the baseline runtime on a set of real-world applications.

\subsubsection*{$N$-body simulation} We previously introduced $N$-body simulation as an example to showcase task-, command- and instruction-graph generation.
While much more efficient approaches exist for the underlying physics problem, the direct $\mathcal{O}(N^2)$ approach exposes an ``all-gather'' data access pattern that is common to many other applications.

\subsubsection*{RSim} We also investigate the iterative radiosity kernel of \textit{RSim}, a room-response simulation for time-of-flight cameras~\cite{celerity-rsim}.
RSim is unique in its growing data access pattern, where each time step kernel appends a new row to a buffer after reading the results of all previous time steps.
This pattern causes frequent allocation resizes unless scheduler lookahead (\cref{sec:lookahead}) is active.

\subsubsection*{WaveSim} The last application we evaluate is \textit{WaveSim}, a five-point wave propagation stencil previously used in other benchmarks of Celerity~\cite{celerity-ccgrid}.
It is computationally inexpensive and only requires a neighborhood data exchange, which makes it a good indicator for executor latency issues.

\subsection{Benchmarking Environment}

All performance data is gathered on the Leonardo supercomputer located in Bologna, Italy, ranking 9th on the TOP500 in November 2024\footnote{\url{https://top500.org/lists/top500/2024/11/}}.
Its Booster Module is a 3456-node cluster of 32-core x86 nodes with 512~GB RAM and 4~NVidia A100 64GB GPUs each, connected via quad-100 Gb/s Infiniband HDR.
The cluster runs Red Hat Enterprise Linux 8.7 with kernel version 4.18, CUDA version 12.1 and OpenMPI 4.1.6.

We build both baseline Celerity and the proposed improved runtime using Clang 17 \texttt{\small -O3} and AdaptiveCpp\footnote{https://github.com/AdaptiveCpp/AdaptiveCpp} v24.06 with the CUDA multi-pass compiler as the SYCL runtime.

Our implementation pins the affinity of main, scheduler, executor and backend threads to dedicated CPU cores in order to avoid jitter in execution time between nodes.

\subsection{Strong-Scaling Behavior}
\label{sec:strong-scaling}

In the context of our research, we are most interested in the runtime's strong-scaling behavior, i.e.\ how speedup on a fixed problem size evolves with the number of participating nodes and GPUs.
This metric helps expose which scheduling inefficiencies the instruction-graph based approach is able to avoid, and what effects limit further scalability.

\begin{figure*}
  \includegraphics[width=0.9\textwidth]{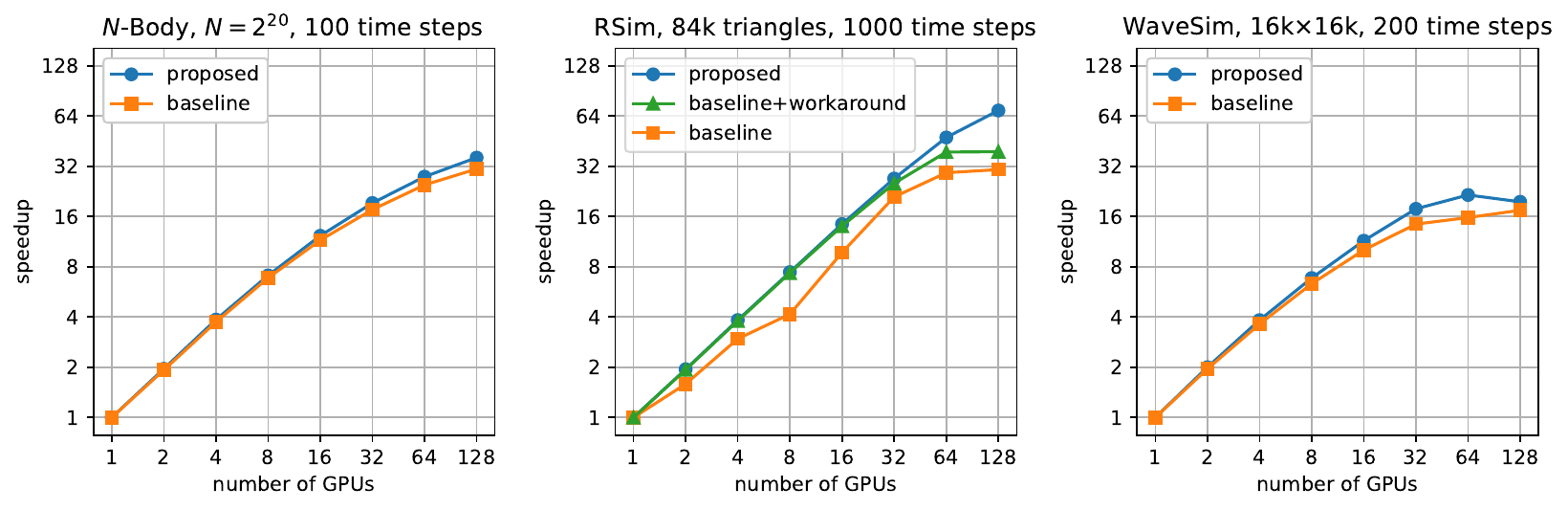}
  \vspace{-0.5em}
  \caption{%
    Strong-scaling behavior of $N$-body simulation, the RSim radiosity kernel, and the WaveSim stencil for fixed problem sizes.
    We compare the existing baseline version of Celerity against the proposed architecture with instruction-graph scheduling.
    One of RSim's performance limitations in the baseline runtime can be lifted using a workaround.
    }
  \label{fig:strong-scaling}
\end{figure*}

\Cref{fig:strong-scaling} shows the speedup of all three test applications for fixed input sizes.
Each node of our test system contains four GPUs, so a run on 128 GPUs is distributed over 32 cluster nodes.
We report the median between 10 benchmark runs.

As is inevitable in strong-scaling experiments, the speedup of a fixed problem size diminishes as the number of processors increases.
However, it becomes apparent that baseline Celerity with ad-hoc memory management (\cref{sec:buffer-manager}) consistently has worse scaling behavior compared to the instruction-graph scheduling runtime.

\subsubsection*{$N$-body strong scaling} The speedup of $N$-body simulation, here performed with 100 time steps on $N=2^{20}$ bodies, diminishes at roughly the same processor count for both implementations.

For this application, the limiting factor is the parallelism exposed by the kernels themselves:
A kernel range of $N=2^{20}$ with a fixed work-group size of 128 means that each of the 128 GPUs is assigned 64 work groups, occupying well below the A100's 108 total streaming multiprocessors.

The small advantage of the instruction-graph runtime is then explained by a more efficient handling of the communications stage between kernels.

\subsubsection*{RSim strong scaling} The unique growing access pattern of RSim makes it an adversarial example for the baseline runtime, triggering an allocation resize in every time step.
This resize can be avoided by submitting a no-op kernel which zero-initializes (and thus allocates) the entire buffer at the start of the program; a workaround which is not feasible to perform by a normal user since it requires an intimate understanding of the runtime's memory management.

We benchmark the radiosity kernel against a scene with 84,000 triangles.
The strong-scaling diagram shows both the naive version and the workaround scaling considerably worse than the proposed IDAG runtime.
RSim has an ``all-gather'' communication pattern comparable to $N$-body, but unlike it, RSim's kernel scales well with the number of GPUs, putting more pressure on the communication logic.

\subsubsection*{WaveSim strong scaling} The 2-dimensional stencil has enough inherent parallelism for kernels to remain efficient even when split between a large number of devices.
As kernel times get shorter, communication and synchronization overhead quickly dominate and expose inefficiencies in scheduling and execution.


\subsection{Scheduling Concurrency and Lookahead}

\begin{figure*}
  \includegraphics[width=\textwidth]{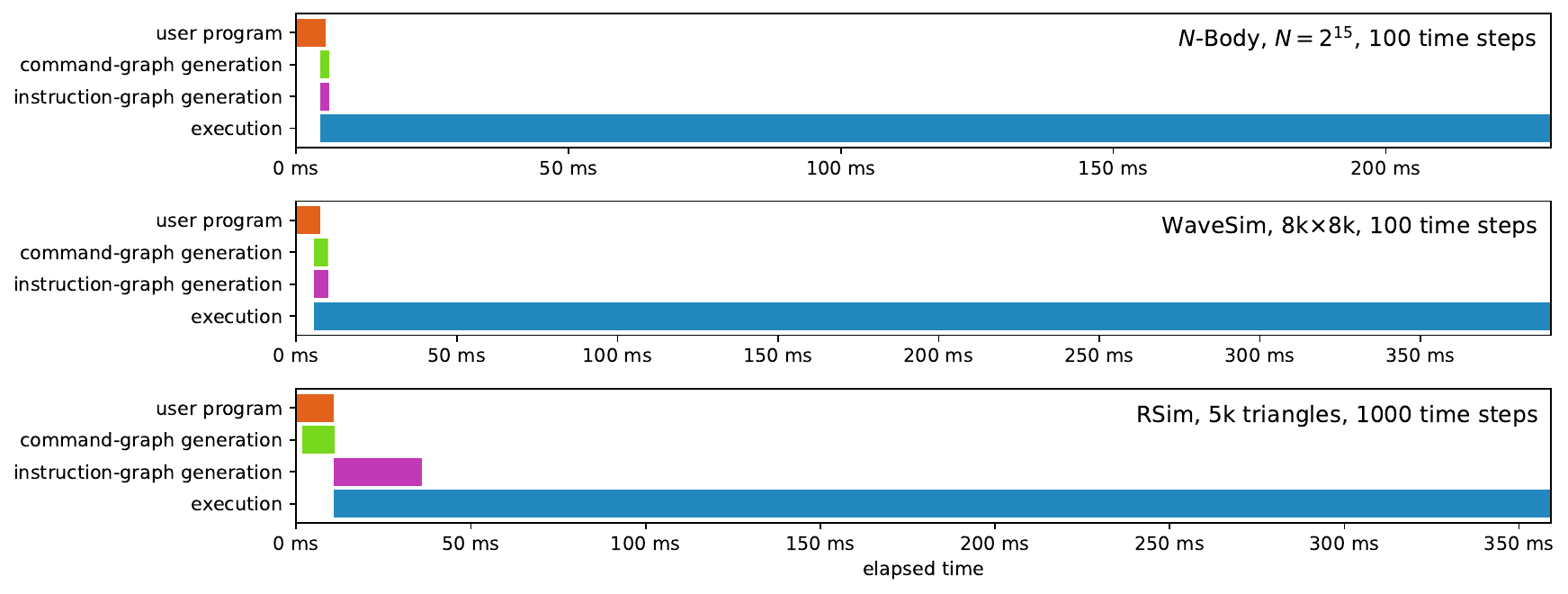}
  \vspace{-1.5em}
  \caption{%
    Timelines for single-node runs of small problem sizes for all three example applications with the proposed architecture.
    Command- and instruction graph generation share the dedicated scheduler thread, and execution is distributed over multiple host threads as well as SYCL queues.
    Both occur asynchronously with the user program.
    }
  \label{fig:profile}
\end{figure*}

The instruction-graph based system architecture effectively decouples scheduling work from execution, allowing both to occur concurrently.
We visualize this behavior in \cref{fig:profile} through runtime profiles of single-node runs on 4~GPUs with small problem sizes for the example applications.

When the user program submits work to Celerity in the form of \textit{command groups}, accompanying task objects are created within the main thread.

Following \cref{fig:architecture}, references to the generated tasks are passed to the scheduler thread, where the command- and instruction graph are generated.
For $N$-body and WaveSim, generation of these graphs is interleaved in order to emit the first executable instruction as soon as possible.

In the RSim case however, scheduler lookahead (\cref{sec:lookahead}) detects that every iteration of the radiosity simulation requires a differently-sized buffer backing allocation, and thus queues commands until the entire command graph has been generated.
Once the instructions for the first iteration are subsequently emitted, they will contain the correctly-sized buffer allocation covering the entire program.

Instruction execution is coordinated by the dedicated executor thread, which distributes work between host threads, backend submission threads, and the GPUs.

The time saved through concurrent scheduling is greatest in the RSim example shown here, where graph generation is relatively time-consuming.
The ratio between scheduling and execution cost can vary drastically depending on the application and the number of participating nodes.

\section{Related Work}



Early works on abstraction layers for GPU-accelerated systems include rCUDA~\cite{rcuda} and GridCuda~\cite{gridcuda}, both of which allow transparent access to remote GPUs by providing wrappers around CUDA API calls.
The former is focused on GPU kernel offloading in heterogeneous clusters, while the latter enables a single node to interface with multiple remote devices.
Both employ a client/server model and do not include facilities for splitting a single kernel across multiple devices.

libWater~\cite{libwater} extends OpenCL's event-based model to express dependencies between computations on a distributed-memory systems.
While memory operations and kernel executions are expressed in terms of specific devices, libWater uses pattern matching on the implicitly constructed dependency graph to substitute point-to-point communication with MPI collective operations.

SnuCL-D~\cite{snucl-d} proposes to execute unmodified OpenCL programs with virtualized remote devices in an SPMD fashion in order to avoid the bottlenecks introduced by a client/server architecture.
The user program and manually distributes work across remove devices, and each node participates only in commands that pertain to either its local devices, or to data that must be sent to or received from remote devices.
Concurrent accesses to the same memory object are serialized to ensure that the location of up-to-date data remains unambiguous.

CUDASA~\cite{cudasa} extends CUDA's hierarchy of thread blocks and grids to include tasks spanning multiple devices, as well jobs spanning clusters of nodes.
Memory is exposed as a single global address space that is shared across all nodes within the cluster.
Allocations are split into evenly sized segments distributed across nodes, which can then be accessed using MPI's one-sided communication primitives.
Since no information about tasks' data requirements is available ahead of time, CUDASA is unable to leverage asynchronous data transfers or locality-aware scheduling.

DistCL~\cite{distcl} combines all devices in a cluster as a single virtual aggregate device, across which it evenly splits OpenCL kernels along work group boundaries.
To perform the required data transfers between nodes, DistCL employs user-defined \textit{meta-functions} which describe kernel memory access patterns.
Performance is sensitive to how fragmented buffer accesses are, with highly fragmented accesses leading to substantial overheads.

SYCL on its own has shown its usefulness as a building block for cluster applications, for example as part of an integration into the HPX runtime~\cite{HPX} through Kokkos~\cite{kokkos} which has been used in astrophysics simulations~\cite{hpx-kokkos-sycl}.

\subsection{Driver-Level Graph Representations}

The efficiency gains from introducing the instruction-graph representation suggest that integrating Celerity with other low-level backend interfaces might help eliminate yet more overhead from the runtime.

As outlined in \cref{sec:sycl-dag}, SYCL itself defines its semantics in terms of dependency graphs, so lowering parts of Celerity's instruction graph to a SYCL DAG is possible.
However, at least AdaptiveCpp and DPC++, being the most popular current SYCL implementations, map their graphs to in-order queues in a manner similar to the concept shown in \cref{sec:architecture}.

Among vendor-specific solutions, Nvidia provides CUDA Graphs as an alternative to its more general stream-based API.
They allow recording a dependency graph of GPU operations up-front and instantiating it to be re-played any number of times.
This helps significantly reduce launch overheads for programs with short-running kernels such as task-parallel or AI workloads~\cite{cuda-speedup-post}.

CUDA Graphs have been used as the basis for cudaFlow, a higher-level framework for constructing GPU task graphs~\cite{cuda-flow}.
They have also been demonstrated as a viable target for transformation of OpenMP-based programs~\cite{openmp-to-cuda-graphs}.


\section{Conclusion}

This paper introduces instruction-graph scheduling as a means of constructing optimized execution plans for multi-GPU distributed parallel programs.
It details how this low-level graph representation can be generated from a command graphs at cluster-node granularity in order to retain maximum concurrency between constituent operations.

We implement our approach in terms of the established Celerity runtime for GPU clusters, creating a system architecture that decouples scheduling from program execution to perform both concurrently.
By limiting responsibility of the executor to efficient instruction dispatch, we demonstrate how a state-machine based logic can assign work to GPUs and OS threads with minimal additional latency.

We further introduce a lookahead scheme on the higher-level command graph representation to reliably avoid costly resize allocations on virtualized buffers, all without adding recurring latency to programs with stable access patterns.

On the basis of three real-world example applications, we demonstrate the superior strong-scaling behavior of the proposed instruction-graph scheduling method compared to the baseline runtime implementation.

Finally, by visualizing profiler timelines on small single-node runs, we show that scheduling is effectively overlapped with execution to varying degrees in all three applications.

\begin{acks}
This research is supported by the Austrian Research Promotion Agency (FFG) via the UMUGUC project (FFG \#4814683).
\end{acks}


\bibliographystyle{ACM-Reference-Format}

\bibliography{paper}

\end{document}